\documentclass[jap,graphicx, reprint, floatfix]{revtex4-1} 
\usepackage{amsfonts, amsmath, graphicx}
\usepackage[version=3]{mhchem}

\begin{document}
\title{Gauge field in systems with spin orbit interactions and additional discrete degrees of freedom to real spin } 
\author{Zhuo Bin Siu}
\affiliation{Computational Nanoelectronics and Nanodevices Laboratory, Electrical and Computer Engineering Department, National University of Singapore, Singapore} 
\author{Mansoor B. A. Jalil} 
\affiliation{Computational Nanoelectronics and Nanodevices Laboratory, Electrical and Computer Engineering Department, National University of Singapore, Singapore} 
\author{Seng Ghee Tan} 
\affiliation{Data Storage Institute, Agency for Science, Technology and Research (A*STAR), Singapore} 

\begin{abstract}
	The spin gauge field formalism has been used to explain the emergence of out of plane spin accumulation in two-dimensional spin orbit interaction (SOI) systems in the presence of an in-plane electric field. The adiabatic alignment of the charge carrier spins to the momentum dependent SOI field, which changes in time due to the electric field, can be mathematically captured by the addition of a gauge term in the Hamiltonian. This gauge term acts like an effective, electric field dependent magnetization. In this work we show that this effective magnetization can be generalized to systems which include additional discrete degrees of freedom to real spin, such as the pseudospin and/or valley degrees of freedom in emerging materials like molybdenum sulphide and silicene. We show that the generalized magnetization recovers key results from the Sundaram-Niu formalism as well as from the Kubo formula. We then use the generalized magnetization to study the exemplary system of a topological insulator thin film system where the presence of both a top as well as a bottom surface provides an additional discrete degree of freedom in addition to the real spin.   
	
\end{abstract} 
\maketitle

\section{Introduction}

	In the Spin Hall Effect (SHE) \cite{SHE1,SHE2,SHE3, SHE4, SHE5}, the passage of an in-plane electric field in a two-dimensional electron gas (2DEG) with spin orbit interactions (SOIs) leads to the appearance of an out of plane spin accumulation. 
	
	Murakami \cite{Murakami} and Fujita \cite{Grp1,Grp2,Grp3,Grp4,SGTSciRep}, and their respective coauthors, had independently studied the SHE. They showed that the out of plane spin accumulation can be understood as the response of the charge carriers as their spins align adiabatically with the momentum dependent SOI field. The direction of the SOI field changes in time due to the change in the momentum of the charge carriers as they accelerate under the electric field. Mathematically, the electric field gives to an effective magnetization term in the Hamiltonian which we shall, for short, call the Murakami-Fujita (MF) potential.
	
	Many emerging material systems in interest in spintronics, for example silicene \cite{Sil1,Sil2,Sil3,Sil4} and  \ce{MoS2} \cite{Mo1, Mo2, Mo3}, possess discrete degrees of freedom (DoFs) such as the pseudospin and / or valley degrees of freedom, in addition to their real spins. In this work, we show in the following sections that the MF potential can be readily extended to incorporate these additional degrees of freedom (DoF) which we shall for simplicity refer to collectively as pseudospin.   To first order in the electric field, the MF potential accounts for the effects of a constant, in-plane electric field for the purposes of calculating spin / charge currents and spin accumulations to first order in the electric field.  
	
	We illustrate the application of the MF potential on a system with a spin$\otimes$pseudospin degrees of freedom through the example of the topological insulator (TI)  thin film system \cite{PRB80_205401,PRB81_041307,PRB81_115407}. Unlike a semi-infinite TI slab, a TI thin film has both a top and a bottom surface which, due t the finite thickness, couple to each other. The low energy effective Hamiltonian for can be written as
\begin{equation}
	H = v(\vec{k}\times\vec{\sigma})\cdot\hat{z} \tau_z + \lambda \tau_x + \vec{M}\cdot\vec{\sigma} \label{TIham1}
\end{equation}
Besides the real spin denoted as $\vec{\sigma}$ of the charge carriers there is another discrete degree of freedom $\vec{\tau}$ associated with whether the charge carriers are localized nearer the upper (  $|+\tau_z\rangle\langle +\tau_z|$ ) or lower ($|-\tau z \rangle\langle -\tau_z|$) surface of the film. The $\tau_x$ term then represents the coupling between the two surfaces of the film due to the finite thickness.    

	This paper is organized as follows. We first revisit the emergence of the MF potential in a spin 1/2 SOI system.  We then generalize the MF potential to include other discrete DoFs, and provide three evidences to support our claim that the MF potential accounts for the effects of the electric field in the sense that an effective Hamiltonian can be constructed by replacing the $\vec{E}\cdot\vec{r}$ term in the original Hamiltonian can be replaced by the position-\textit{independent} MF potential. 
	
	We first show that taking the momentum derivative of the effective Hamiltonian in the Heisenberg equation of motion for the position operator reproduces the usual Berry curvature expression for the anomalous Hall velocity. 	As part of their paper on the microscopic origin of spin torque \cite{ChengRan}, Cheng Ran and Niu Qian had extended the original Sundaram-Niu wavepacket formalism \cite{ChioGoo1}, which gave only the time variation of the position and momentum expectation values, to now include the time variation of the spin expectation values. We show that Ran and Niu's expressions for the time evolution of the spin expectation values can be readily extended to incorporate the other discrete degrees of freedom present, and that the time evolution of these operators can be derived from applying the Heisenberg equation of motion to the MF potential. Finally, we show that the Kubo expression for non equilibrium expectation of spin$\otimes$pseudospin quantities can be interpreted as the first order time independent perturbation theory response to the MF potential. 
	
	We then move on to apply the MF potential formalism to study the emergence of a TI thin film system subjected to an in-plane magnetization and electric field. We first illustrate the effects of the interlayer coupling on the in-plane magnetization and the dispersion relations. We then show that the direction of the out of plane spin accumulation resulting from an in-plane electric field can be explained in terms of how the direction of the momentum dependent in-plane SOI field rotates with the change in momentum direction resulting from the electric field. The anti-symmetry of the out of plane spin accumulation in $k$ space can be broken with the application of an out of plane electric field in order to yield a finite spin accumulation after integrating over the Fermi surface.  

\section{Spin 1/2 systems}\label{spinhalf}
	To familarize the reader with the MF potential, we first review its appearance in spin 1/2 SOI systems without any additional discrete degrees of freedom. The Hamiltonian for a homogenous 2DEG with SOI and an electric field $E_x$ in the $x$ direction can be generically written as
\[	H = \frac{p^2}{2m} + \vec{B}(\vec{k})\cdot\vec{\sigma} + E_x x 
\]
where the $\vec{B}(\vec{k})$ represents a momentum dependent spin orbit interaction. We define a unitary transformation $U(\vec{k})$ which diagonalizes $\vec{B}.\vec{\sigma}$ in spin space so that after the unitary transformation, we have 
 \[
 	UHU^\dagger = \frac{p^2}{2m} + |\vec{B}|\sigma_z + E_x (x - i U \partial_{k_x} U^\dagger).
 \]

Mathematically, the effect of $U$ can be interpreted as rotating the spin space coordinates so that in the rotated frame, the spin $z$ axis points in the direction of the SOI field $\vec{B}(\vec{k})$.  The non commutation between $x$ and the momentum dependent $U$ results in the appearance of the $-i E_x (U \partial_{k_x} U^\dagger)$ term which acts as an effective magnetization $M'_i \tilde{\sigma}_i$ in the rotated frame where the tilde on the $\tilde{\sigma}_i$ indicates that the index $i$ refers to the $i$th spin direction in the rotated frame. To determine what \textit{lab} frame direction this effective magnetization points in, we perform the inverse unitary transformation $U^\dagger (-i U \partial_{k_x}U^\dagger) U = -i (\partial_{k_x}U^\dagger)U$. This expression can be evaluated without an explicit form for $U$. To do this, we first note that by definition $U\hat{b}\cdot\vec{\sigma}U^\dagger = \sigma_z$ where $\hat{b} = \vec{B}/|\vec{B}|$. Thus,
\begin{eqnarray*}
	&& U^\dagger\sigma_zU = \hat{b}\cdot\vec{\sigma} \\
	&\Rightarrow& (\partial_{k_x}U^\dagger)\sigma_z U + U^\dagger\sigma_z(\partial_{k_x}U) = \partial_{k_x}\hat{b}\cdot\vec{\sigma} \\
	&\Rightarrow& [(\partial_{k_x}U^\dagger)U, \hat{b}\cdot\vec{\sigma}]  = \partial_{k_x}\hat{b}\cdot\vec{\sigma} \\
\end{eqnarray*}
In going from the first to second line, we differentiated the first line with respect to $k_x$ and then inserted $\mathbb{I}_\sigma=UU^\dagger$ in the appropriate places. From the last line, we use the fact that $[\vec{a}\cdot\vec{\sigma}, \vec{b}\cdot\vec{\sigma}] = i (\vec{a}\times\vec{b})\cdot\vec{\sigma}$ to conclude that \[
	-i E_x U\partial_{k_x}U^\dagger = E_x (\hat{b}\times\partial_{k_x}\hat{b})\cdot\vec{\sigma}. 
\] 

This is the MF potential for a spin 1/2 system with an electric field in the $x$ direction. 

Notice that although $U$ is not unique,  the lab frame direction of $-i U\partial_{k_x}U^\dagger$ is independent of the specific choice of $U$. $E_x(-i U\partial_{k_x}U^\dagger)$ can then be thought as as an electric field dependent effective magnetization which confers a spin accumulation in the $(\hat{b}\times\partial_{k_x}\hat{b})\cdot\vec{\sigma}$ direction. Taking the specific example of the Rashba SOI where $\vec{B} = \alpha( p_y, -p_x)$ both $\vec{B}$ and $\partial_{k_x}\vec{B}$ lie on the $xy$ plane.  $E_x(\hat{b}\times\partial_{k_x}\hat{b})$ thus points in the out of plane spin direction, and confers an out of plane spin accumulation to the charge carriers 

\begin{figure}[ht!]
\centering
\includegraphics[scale=0.5]{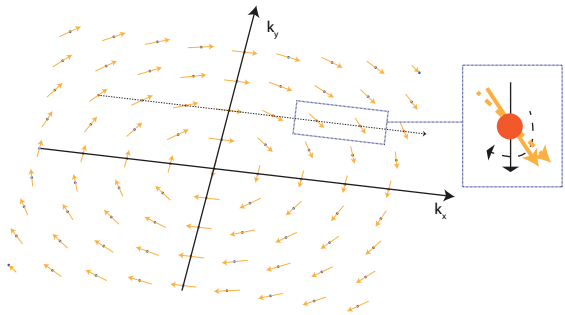}
\caption{ The arrows at each point in $k$ space indicate the direction of the Rashba SOI field there. The application of the electric field causes the momentum of the charge carrier to trace out the trajectory in $k$ space indicated by the dotted line. The spin of the charge carrier adiabatically follows the direction of the SOI field at each point in $k$ space. The rotation of the spin can be thought of as being due to an effective magnetization which both creates the torque necessary to rotate the spin as shown in the inset, as well as confers an out of plane spin accumulation. } 
\label{gA2SOIfield} 
\end{figure}

Physically, the origin of the $(\hat{b}\times(E_x \partial_{k_x}\hat{b})$ term can be explained by assuming that the spins of the charge carriers adiabatically follow the direction of the SOI field. As shown in Fig. \ref{gA2SOIfield}, $\vec{B}(\vec{k})\cdot\vec{\sigma}$ associates each point in $k$ space with a SOI field pointing in the $\hat{b}(\vec{k})$ direction. Assume that the electric field is initially switched off and consider a carrier with a definite $\vec{k}$. As the electric field is switched on, the field causes the charge carrier to accelerate in the direction of the field so that the momentum changes and the carrier traces out a trajectory along $k$ space. We assume that the electric field is weak enough so that the spin of the carrier rotates along with the direction of the SOI field as it successively moves through different $k$ points. The resulting rotation of the spin can be thought of as being due to an effective magnetic field pointing along the $\hat{b}\times\partial_t \hat{b} =  E_x \hat{b}\times\partial_{k_x} \hat{b}$ direction which both provides the torque necessary to rotate the spin as well as confers a spin accumulation in the direction of the torque. 

We now proceed to a general description of the MF potential generalized to include other discrete degrees of freedom.

\section{The Murakami-Fujita potential} 
Consider now a generic Hamiltonian
\begin{equation}
	H_0  = b_i(\vec{k})\kappa_i \label{H0}  
\end{equation}	
where the $\kappa_i$s are finite sized matrices representing the discrete degrees  of freedom. For example, in a spin 1/2 system with SOC, the 4 $\kappa_i$s are the Pauli matrices and the identity matrix. In the TI thin film Hamiltonian Eq. \ref{TIham1} the $\kappa_i$s represent the 16  $\vec{\sigma}\otimes\vec{\tau}$ matrices. 

In order to write the Hamiltonian Eq. \ref{H0} as a numerical matrix, we need to express the matrix elements in terms of basis. For example, for spin 1/2 system it is common to adopt the usual representation of the Pauli matrices so that, for instance $H_0 = \vec{b}\cdot\vec{\sigma} \simeq \begin{pmatrix} b_z & b_x - i b_y \\ b_x + i b_y & -b_z \end{pmatrix}$.  The numerical matrix on the rightmost side of the equal sign is written in the $|\pm z \rangle$ basis. We refer to the basis which $H_0$ as a `numerical matrix' is in as the `laboratory frame' with basis states $|\lambda_i\rangle$ ($\lambda$ for \textit{l}aboratory. ) Label now the $i$th eigenstates of $H_0$ by $|\epsilon_i\rangle$. We assume that the laboratory basis is fixed, i.e. it has no dependence on any parameter in the Hamiltonian so that, for instance $\partial_{k_x} |\lambda_i\rangle = 0$. Instead of using the laboratory basis, we can also expand our states and operators in terms of the eigenbasis, and convert between the two basis through the unitary transformation $U$. Defining the $U$s so that $UH_0 U^\dagger$ is diagonal in the eigenbasis representation, we have 
 \begin{equation}
 	U = \sum_{i,j}  |\epsilon_i \rangle \langle \epsilon_i|\lambda _j\rangle \langle j|
 \end{equation},
i.e. the matrix $i,j$th  elements in the numerical representation of $U$ is $\langle \epsilon_i|\lambda_j\rangle$. Notice that since the phase factor  $\exp(i\phi)$ can be introduced to $|\tilde{\epsilon}_a \rangle = |\epsilon_i\rangle\exp(i\phi_a)$ arbitrarily the values of the matrix elements $\langle  \tilde{\epsilon}_i | \lambda_j \rangle$ will vary with the phase of $|\epsilon_i\rangle$s. Now consider adding a perturbative electric field .  The Hamiltonian then becomes $H = H_0 + E_x x$, and we have  $UHU^\dagger = UH_0U^\dagger + E_x ( x + i U\partial_{k_x} U^\dagger)$ where, in this rotated frame, $H_0$ is diagonal, and we have an additional $i U\partial_{k_x}U^\dagger$. In order to figure out the lab frame spin$\otimes$pseudospin `direction' where this contribution points to, we transform the $i U\partial_{k_x}U^\dagger$ piece  \textit{without the diagonal elements} back to the laboratory frame . The reason for the removal of the diagonal elements will become apparent later. With the diagonal elements in place, we have  $U^\dagger (i U\partial_k U^\dagger) U = -i U^\dagger\partial_k U$. (We have dropped the suffix $x$ from $k_x$ and $E_x$ for notational simplicity)

We stress that $-i U^\dagger\partial_k U$ has the same numerical matrix elements in the laboratory frame regardless of the  phases of the $\langle \lambda_i | \epsilon_j \rangle$. This is because

\begin{eqnarray*}
	&& -i U^\dagger \partial_k U  \\
	&=& -i |\lambda_a \rangle \langle \lambda_a|\epsilon_b \rangle \langle \partial_k \epsilon_b|\lambda_c \rangle \langle \lambda_c | \\
	&=& -i |\epsilon_a \rangle \langle \partial_k \epsilon_a|	
\end{eqnarray*}
The second line gives the numerical values of the laboratory frame $ac$th matrix elements, and the third line the simplification using a resolution of identity. 
Notice that we have the combination $|\epsilon_b \rangle \langle \partial_k \epsilon_b|$ with the same state index $b$ occurring  together so that any phase factor $\exp(i \phi_b)$ introduced in $|\epsilon_b\rangle \rightarrow |\epsilon_b \rangle\exp(i \phi_b)$ cancels out. 

Returning now to the diagonal elements of the rotated frame $i U\partial_k U^\dagger$, we see that they correspond to $i |\epsilon_i \rangle \langle \epsilon_i|\partial_k |\epsilon_i\rangle\langle \epsilon_i|$.  Subtracting them off from $-i U^\dagger\partial_k U$ gives the MF potential $H_{MF}$ where 
\begin{equation}
	H_{MF} = -i \sum_{a \neq b} |\epsilon_a \rangle \langle \partial_k \epsilon_a|\epsilon_b \rangle \langle \epsilon_b| E.
\end{equation}

We argue that, at least for the purposes of calculating currents and spin$\otimes$pseudospin accumulations the effects of the electric field $E_i$ to the first order in $E$ can be incorporated by replacing $E_i x_i$ with $H_{MF}$ so that the effective Hamiltonian reads
\begin{equation}
	H' = H_0 + H_{MF} = b_i(\vec{k})\kappa_i - i \sum_{a \neq b} |\epsilon_a \rangle \langle \partial_{k_i} \epsilon_a|\epsilon_b \rangle \langle \epsilon_b| {E_i}. \label{Heff}
\end{equation}

In order to support our claim, we list three examples where the use of $H'$ recovers well-known results. 
 
\subsection{Anomalous velocity}
In the presence of an electric field, charge carriers can acquire an anomalous velocity \cite{AV1,AV2, AV3} proportional to the Berry curvature \cite{AV4, AV5, ChioGoo1, ChioGooRev}. We show that this result can be recovered via the Heisenberg equation of motion on Eq. \ref{Heff}.    Under the Heisenberg equation of motion, we have $\partial_t \vec{r} = -i  ( [\vec{r}, H_0]   + [\vec{r}, H_{MF}] = \nabla_{\vec{k}} (H_0 + H_{MF}) $ The first term is the usual velocity. We shall show that the second gives the usual Berry curvature anomalous contribution to the velocity. Taking the expectation value of the second with respect to the $i$th eigenstate of $H_0$, we have 
\begin{eqnarray*}
	&& -i \langle \epsilon_i | [x_b, H_{MF}]  |\epsilon_i \rangle \\
	&=& -i E_b \langle \epsilon_i| \partial_{k_b} \Big( \sum_{\substack{jk \\ j\neq k}} |\epsilon_j \rangle \langle  \partial_{k_a}\epsilon_j| \epsilon_k \rangle\langle\epsilon_k| \Big) |\epsilon_i \rangle \\
	&=&-i E_b \sum_j (\langle \epsilon_i |\partial_{k_b} \epsilon_j \rangle \langle \partial_{k_a}\epsilon_j|\epsilon_i\rangle + \langle \partial_{k_a} \epsilon_i | \epsilon_j\rangle\langle \partial_{k_b} \epsilon_j|\epsilon_i \rangle) \\
	&=& 2 E_b \mathrm{Im} ( \langle \partial_{k_a} \epsilon_i|\partial_{k_b} \epsilon_i \rangle).
\end{eqnarray*}

The third line is simply the expansion of the $\partial_{k_b}$ differential. Notice that the requirement that $j \neq k$ stemming  from the removal of the diagonal terms of the rotated frame  $-i U^\dagger \partial_k U$ results in the absence of terms $\langle \partial_{k_a}\partial_{k_b} \epsilon_i|\epsilon_i\rangle$  and $\langle \partial_{k_b} \epsilon_i|\partial_{k_a} \epsilon_i\rangle$ due to the $\partial_{k_b}$ acting on the second and third terms in the big bracket in the second line. The last line is the usual Berry curvature term for the anomalous velocity. 

\subsection{Spin and other discrete DoFs}
As part of their paper on explaining the microscopic origin of spin torque, Cheng and Niu extended the original Sundaram-Niu formalism, which described only the spatial evolution of position and velocity, to now cover the time evolution of spin 1/2 as well. Their formalism can be easily extended to cover the time evolution of operators with finite discrete spectra. We describe the extension in the appendix, and simply state the end result here.

For a state 
\[
	|\psi\rangle = \sum_i |\epsilon_i \rangle \eta_i 
\]
where the summation $i$ runs over the discrete DoFs (e.g spin up / down for spin 1/2, and the upper / lower surfaces for a TI thin film)  and the continuous quantum numbers (e.g. $\vec{k}$ in SOI systems) and the $\eta_i$s are the weightages of the $i$th basis state, we show in the appendix that for an operator $O$ in the discrete DoFs that 

\begin{equation}
	d_t \langle \psi | O | \psi \rangle = 2E_a  \mathrm{Re} ( \eta_i^* \langle \epsilon_i|\partial_{k_a} \epsilon_j \rangle \langle \epsilon_j |O|\epsilon_k \rangle \eta_k 
\end{equation}

It is straightforward to show that this expression is $-i \langle \psi [ O, H_{MF}] |\psi\rangle$.  

\subsection{Recovery of the Kubo formula}
Treating  $H_{MF}$ as a perturbation to $H_0$ and applying the standard non-degenerate time-independent perturbation theory to the $i$th eigenstate of $H_0$, $|\epsilon_i\rangle$, the first order correction to $|\epsilon_i\rangle$ which we denote as $|\epsilon_i^{(1)}\rangle$ reads 
\[	|\epsilon_i^{(1)}\rangle = \sum_j |\epsilon_j \rangle \frac{ \langle \epsilon_j|H_{MF}|\epsilon_i\rangle}{E_i - E_j}
\]
so that to  the correction to the expectation value of an observable $O$ for state $|\epsilon_i\rangle$ to first order in $\vec{E}$, $\delta \langle i| O | i \rangle$ is 
\begin{eqnarray}
	&& \delta \langle i| O|i \rangle  \\ \nonumber
	&=& 2\mathrm{Re} (\langle \epsilon_i|O|\epsilon_i^{(1)}\rangle ) \\ \nonumber
	&=& 2\mathrm{Re} \sum_j \frac{  \langle \epsilon_i |O| \epsilon_j \rangle \langle \epsilon_j |H_{MF}|\epsilon_i\rangle}{E_i - E_j}. \label{c1oe}
\end{eqnarray}

However,  since 
\begin{eqnarray*}
	&& \partial_k \langle \epsilon_i | H_0 |\epsilon_i \rangle = 0 \\
	&\Rightarrow& \langle \partial_k \epsilon_i |\epsilon_j \rangle (E_i-E_j) = \langle \epsilon_i|\partial_k H_0 |\epsilon_j  \rangle \\
	&\Rightarrow& \langle \partial_k \epsilon_i|\epsilon_j \rangle = \frac{ \langle \epsilon_i| \partial_k H_0 |\epsilon_j \rangle}{E_i - E_j},
\end{eqnarray*}
 we can rewrite 
\begin{eqnarray*}
	H_{MF} &=& -i E_i \sum_{a \neq b} |\epsilon_a \rangle \langle \partial_{k_i} \epsilon_a|\epsilon_b \rangle \langle \epsilon_b|. \\
	&=&-i \sum_{a \neq b} |\epsilon_a\rangle  \frac{ \langle \epsilon_a| \partial_{k_i} H_0 |\epsilon_b \rangle}{E_a - E_b} \langle \epsilon_b|.
\end{eqnarray*}

A common form of the Kubo formula is 
\begin{equation}
	\delta \langle O \rangle \propto \sum_{\vec{k}} \sum_{a= \neq  b} \frac{n(E_a)-n(E_b)}{(E_a-E_b)^2} \mathrm{Im} ( \langle a|O|b\rangle \langle b| (\partial_k H_0)|a \rangle \label{kuboPaper}. 
\end{equation}

Substituting this back into Eq. \ref{c1oe} gives a result similar to the Kubo expression for the change in an expectation value of $O$ under an electric field -- 
\begin{equation}
	\delta \langle i| O|i \rangle  = 2 \mathrm{Im} \sum_j \vec{E}\cdot\frac{  \langle \epsilon_i |O| \epsilon_j \rangle \langle \epsilon_j |\nabla_{\vec{k}} H_0 |\epsilon_i\rangle}{(E_i - E_j)^2}. \label{c2oe}
\end{equation} 

Our result Eq. \ref{c2oe} corresponds to Eq. \ref{kuboPaper} with the occupancy factor $n$ set to 1 for the $i$th state we are interested in and 0 for the other states, and without a second summation over all states.   

Having established the link between the MF potential and the Kubo formula, we now proceed to use Eq. \ref{c2oe} to study the exemplary system of a topological insulator thin film system. 

\section{TI thin films} 

The effective Hamiltonian for the surface states of a TI thin film of infinite dimensions along the $x$ and $y$ directions, and small finite thickness along the $z$ direction, can be written as 
\begin{equation}
		H = (\vec{k}\times\vec{\sigma})\cdot\hat{z}\tau_z + M_y\sigma_y + \lambda \tau_x. \label{tfHam0}
\end{equation}

We use units where $e=\hbar=v_f=1$. 

We first highlight the influence of the inter-surface coupling term $\lambda$ on the energy spectrum. Consider the limit where $\lambda \rightarrow 0$,$M_y \neq 0$. In this case, the upper and lower surfaces may be considered separately, and the energy spectrum consists of two Dirac cones. The states localized near the upper surface have $\langle \tau_z \rangle = +1$, while the state localized near the lower surface have the opposite sign of $\langle \tau_z \rangle$. The $M_y\sigma_y$ term, however, has the same sign for both the upper and lower surfaces. The Dirac points for the Dirac cones for the upper surface states and the lower surface states are hence displaced in opposite directions in $k$ space. 

\begin{figure}[ht!]
\centering
\includegraphics[scale=0.27]{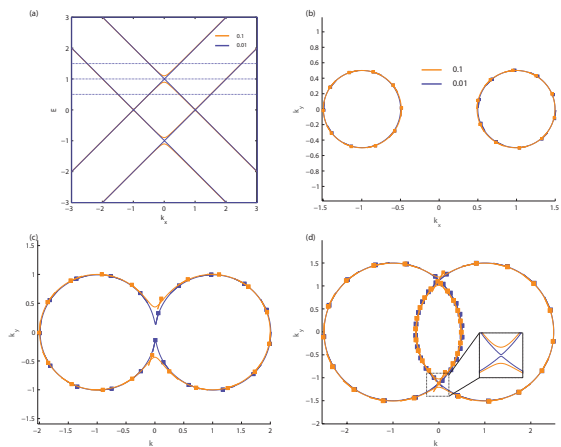}
\caption{ Panel (a) shows the dispersion relations for the two values of $\lambda$ indicated in the legend at $k_y = 0$ for $\vec{M} = 0.5\hat{y}$. The three horizontal dotted lines correspond to the values of energies at which the EECs in panels (b) to (d) at $E = 0.5,\ 1$ and $1.5$ are plotted respectively. Panels (c) and (d) show the EECs and the in-plane spin accumulation directions at the two values of $\lambda$ indicated by different colors in panel (b). The inset of panel (d) shows a zoomed in view of the EECs near the intersection of the two Fermi `circles' showing that the inter surface coupling leads to a breaking away of the lens shaped region where the two circles overlap into separate EEC curves.}
\label{glambdaComb2}
\end{figure}

We now turn on the inter-surface coupling. Fig. \ref{glambdaComb2} shows the dispersion relations and equal energy contours (EECs) at two values of $\lambda < |\vec{M}|$. At these small (relative to $|\vec{M}|$) values of $\lambda$ the two Dirac cones corresponding to the surface states localized at the upper and lower surfaces of the thin film are still distinctly evident. At low values of energy where the two cones do not overlap (panel (b) ), the EECs consist of two almost circular curves that correspond to the cross sections of the two Dirac cones. As the energy increases and the two almost-circular cross sections begin to almost touch each other, the inter-surface coupling pushes the EECs outwards in $k$-space so that the cross sections link up with each other and form a single curve ( panel (c) ). A further increase in energy causes the the two Dirac cones overlap with each other the anti crossing of the energy levels due to the inter surface coupling causes the $k$ space lens-shaped region where the Dirac cones overlap to break away from the outer perimeter of the overlapping `circles' and form a second closed curve. Despite the distortions of the EECs from the perfectly circular profiles in the absence of inter-surface coupling, the directions of the in-plane spin accumulation along the EECs in the presence of inter-surface coupling still roughly follow those of the original Dirac cones. Returning now to panel (a) of the figure, it is evident that as the inter-surface coupling increases, the energy of the lowest energy particle (hole) band at $\vec{k}=0$ increases (decreases).

\begin{figure}[ht!]
\centering
\includegraphics[scale=0.27]{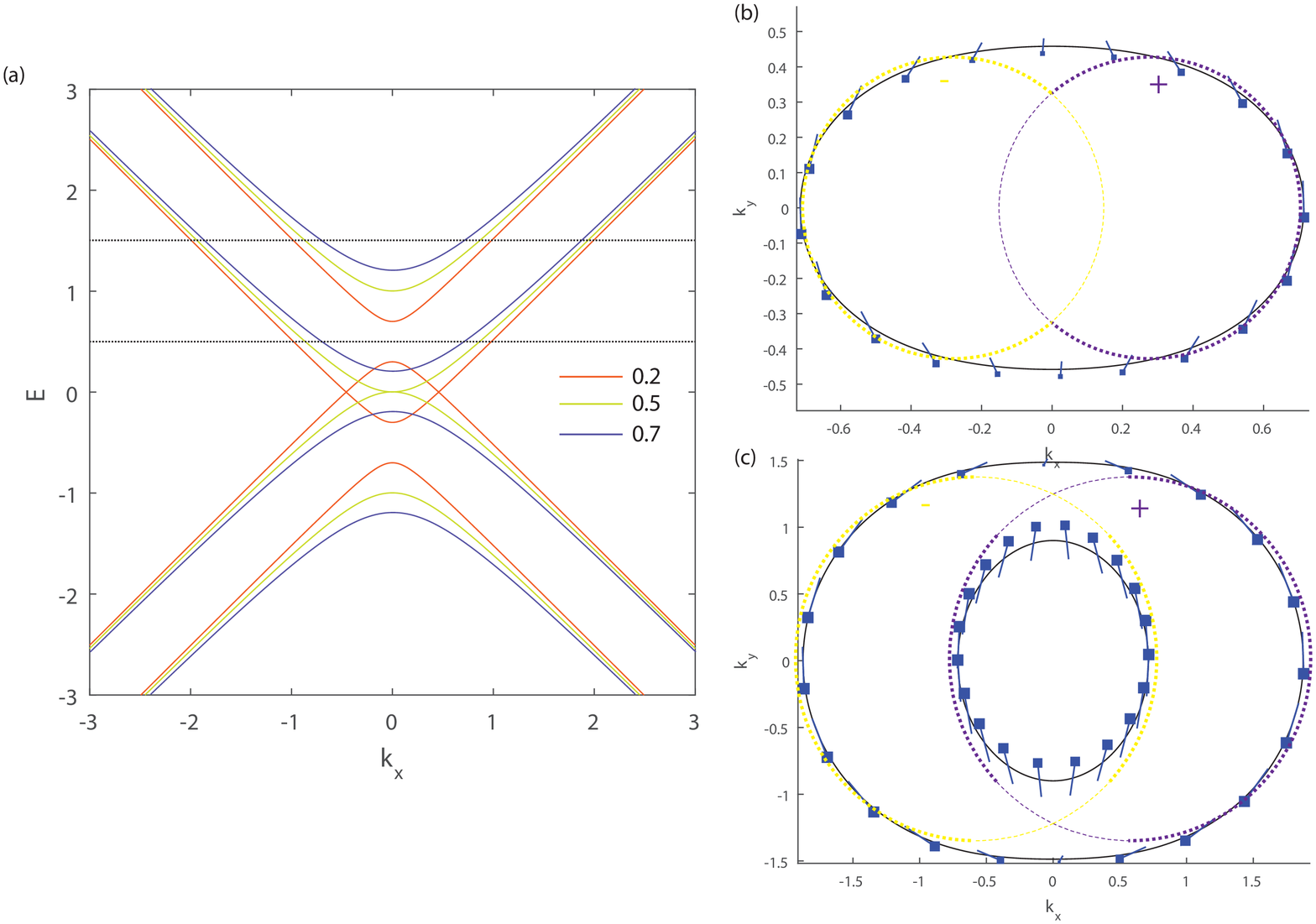}
\caption{ Panel (a) shows the dispersion relations for progressively larger values of $\lambda$ indicated by the colors in the legend relative to the fixed value of $\vec{M} = 0.5\hat{y}$. Panels (b) and (c) show the EECs (in solid lines) and in-plane spin accumulation directions for $\lambda = 0.7$ at the two values of energies (0.5 and 1.5 respectively) indicated by the horizontal dotted liens in panel (a). The two dotted circles in panels (b) and (c) are indicative of the Fermi circles for $\pm +(\vec{k}\times\vec{\sigma})\cdot\hat{z}$ Dirac cones which provide rough indications of the in-plane spin accumulation directions at the $k$ space points on the EECs. }
\label{gbigLambdaComb1} 
\end{figure}

Fig. \ref{gbigLambdaComb1} shows the dispersion relations and the EECs as $\lambda$ increases further relative to $|\vec{M}|$. As $\lambda$ is increased from 0, the energy of the lowest energy particle band at $\vec{k}=0$ is pushed downwards and that of the highest energy hole band pushed upwards until the two bands touch each other when $\lambda = |\vec{M}|$. At this point ($\lambda=0.5$ in panel (a)) we no longer have two the well-resolved Dirac cones with two separate Dirac points in $\lambda = 0.2$ in panel (a) of the figure. A further increase in $\lambda$ leads to a bandgap opening up between the particle and hole bands. Panels (b) and (c) of the figure show the EECs at two values of energy for $\lambda > |\vec{M}|$. The in-plane spin accumulations at various $k$ space points can still be roughly understood as the spin accumulations of two overlapping circular cross sections of perfect Dirac cones. 

\begin{figure}[ht!]
\centering
\includegraphics[scale=0.3]{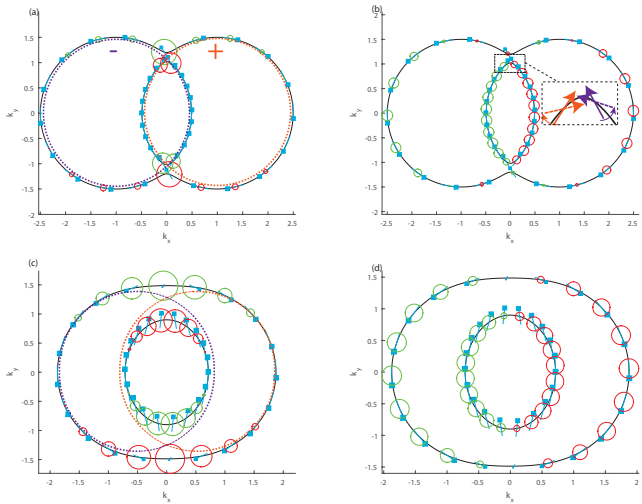}
\caption{ The EECs, in-plane spin accumulation directions and out of plane spin accumulation at two representative values of energy in the $\lambda < |\vec{M}|$ regime  ( (a) and (b) ) and $\lambda > |\vec{M}|$ regime ( (c) and (d) ) for electric fields applied in the $x$ ( (a) and (c) ) and $y$ ( (b) and (d) ) directions. The sizes on the circles on the EECs are indicative of the magnitudes of the out of plane spin accumulations due to the electric field (the sizes of the circles are \textit{not} scaled linearly to the spin accumulation magnitudes) with green (red) circles indicating out of plane spin accumulations in the negative (positive) $z$ directions. The two dotted circles in the left panels are indicative of the Fermi circles for $\pm +(\vec{k}\times\vec{\sigma})\cdot\hat{z}$ Dirac cones which provide rough indications of the in-plane spin accumulation directions at the $k$ space points on the EECs. $E=1.5R$ for all panels; $\lambda = 0.1, \vec{M} = \hat{y}$ for (a) and (b) and $\lambda = 0.7, \vec{M} = 0.5 \hat{y}$ for (c) and (d).   }
\label{gspinZpoln} 
\end{figure}		

We now turn our attention to the out of plane spin accumulation generated by an electrical field which we calculate using Eq. \ref{c2oe} . Fig. \ref{gspinZpoln} shows the out of plane spin $z$ accumulation generated at various $k$ space points on the EECs of a TI thin film with $\lambda > |\vec{M}|$ (panels (a) and (b) ), and $\lambda > |\vec{M}|$ (panels (c) and (d) ) for electrical fields applied in the $x$  ( panels (a) and (c) ) direction perpendicular to the magnetization, and the $y$ direction ( panels (c) and (d) ) parallel to the magnetization.  The sign of the resulting spin $z$ accumulation can be understood in terms of how the applied electric field changes the direction of the SOI field experienced by the charge carriers. We noted in our earlier discussion in Sect. \ref{spinhalf} that each point on the EECs may be associated with the Fermi circle of either the $+(\vec{k}\times\vec{\sigma})\cdot\hat{z}$ Dirac cone, or the $-(\vec{k}\times\vec{\sigma})\cdot\hat{z}$ cone. This is also indicated on the left panels of Fig. \ref{gspinZpoln} where the two Fermi circles are indicated by dotted circles of different colors. Consider now the $k$ space region denoted in the inset of panel (b). The inset shows the spin accumulations on two points in $k$ space with the red (blue) arrows denoting the spin accumulation direction for a point on the + (-) Fermi circle. The passage of an electric field in the $y$ direction causes $\langle p_y \rangle$ to increase while $\langle p_x \rangle$ remains constant, so that the SOI field $\pm (\vec{k}\times\hat{z})$ as well as the spin accumulation rotates in opposite directions for the $\pm$ Fermi circles. Reminiscent of our earlier discussion on spin 1/2 systems, this rotation in turn indicates the existence of an out of plane effective magnetic field which in turn imparts an out of plane spin accumulation. Applying the same argument to most of the other $k$-space points on  the EECs in the figure explains the \textit{sign} of the out of plane spin accumulation there. The \textit{magnitude} of the spin $z$ accumulation depends on how much relative change in the SOI field direction the application of the electric field leads to. For example, in the right panels of the figure, the largest spin $z$ accumulation are on those EEC segments where the in-plane spin accumulation are in the $\pm y$ directions so that the small increment in the SOI field in the $\pm x$ directions due to the $y$ electric field is a large increment compared to other $k$ space points on the EECs where the spin accumulations already have large $x$ components. 

The out of plane spin $z$ accumulations in the preceding figures are antisymetrically distributed in $k$ space on the EECs. This antisymmetry results there being no net out of plane spin accumulation in $k$ space after integrating over the entire Fermi surface. In order to break the antisymmetry, we now introduce a term $E_z \tau_z$ to the Hamiltonian Eq. \ref{tfHam0} so that the Hamiltonian now reads 
\begin{equation}
		H = (\vec{k}\times\vec{\sigma})\cdot\hat{z}\tau_z + \vec{M}\cdot\vec{\sigma} + \lambda \tau_x + E_z\tau_z. \label{tfHam1}
\end{equation} 

The $E_z\tau_z$ term introduces an asymmetry between the upper and lower surfaces. This asymmetry may physically result from the fact that in experimentally grown TI thin films the bottom surface of the film is in contact with the usually non-ferromagnetic substrate, and the upper surface either in contact with the vacuum (for $\vec{M}\cdot\vec{\sigma}$ being due to magnetic doping \cite{PRL102_156603,Sci329_659,NatPhy7_32} ) or with a FM layer (for $\vec{M}\cdot\vec{\sigma}$ being due to the proximity effect with a FM layer \cite{PRL104_146802,PRL110_186807, PRB88_081407} ) .

\begin{figure}[ht!]
\centering
\includegraphics[scale=0.3]{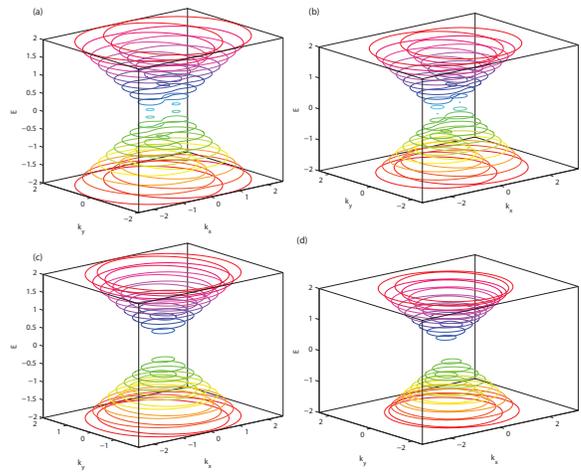}
\caption{ Panel (a) and (b) show the EECs at various energies in the (a) absence and (b) presence of $E_z$ for $\lambda < |\vec{M}|$. Panels (c) and (d) show the EECs at various energies in the (c) absence and (d) presence of $E_z$ for $\lambda > |\vec{M}|$.   ( $\vec{M} = 0.5\ \hat{y}$, $\lambda = 0.2$ in (a) and (b); $\vec{M} = 0.2 \hat{y}$ in (c) and (d). $E_z = 0.1$ in (b) and (d).  )  }
\label{gEZeec} 
\end{figure}		

Fig. \ref{gEZeec} compares the EECs in both the $\lambda < |\vec{M}|$ regime as well as the $\lambda > |\vec{M}|$ regimes in the presence and absence of the $E_z \tau_z$ term. The asymmetry between the upper and lower surfaces of the TI film due to the $E_z\tau_z$ term results in the states stemming from the Dirac cones corresponding to the two surfaces being shifted in opposite directions along the energy axis. The dispersion relations become `tilted', and the EECs at a given value of energy becoming asymmetrical in $k$ space. 

\begin{figure}[ht!]
\centering
\includegraphics[scale=0.3]{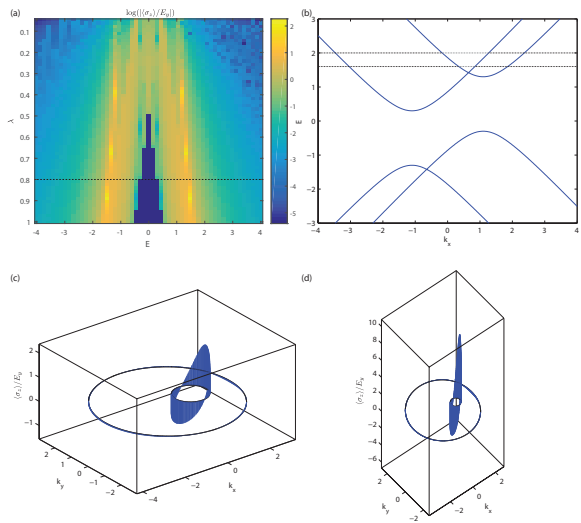}
\caption{ Panel (a) shows the $\log |\langle \sigma_z(E) \rangle|$, the logarithm of the out of plane spin accumulation integrated over the EEC at energy $E$ as a function of $E$ and $\lambda$ at $E_z = 1.1$ and $\vec{M} = 0.5\hat{y}$ for an electrical field in the $y$ direction.  The dotted line in panel (a) corresponds to the value of $\lambda$ in panel (b), which shows the dispersion relation at $k_y = 0$ and $\lambda = 0.8$. The `tilting' of the dispersion relations due to $E_z$ is evident from the plot. The two dotted lines in panel (b) in turn correspond to the energies for which the EECs and the out of plane accumulation at each $k$ space point is plotted in panel (c) for $E = 2$, and (d) for $E = 1.7$ respectively. }
\label{gjyzEz} 
\end{figure}

This asymmetry then results in a net out of plane spin accumulation after integrating over all the $k$ space points spanned by the EECs. Evidently, the spin accumulation increases with the magnitude of $\vec{M}$ and $E_z$. What is perhaps more interesting is the variation of $\langle \sigma_z (E) \rangle $, the out of plane spin accumulation integrated over the EECs at a given value of $E$, with the inter surface coupling $\lambda$ for a fixed value of $\vec{M}$ and $E_z$. Fig. \ref{gjyzEz} shows  

Panel (a) of Fig. \ref{gjyzEz} shows the (\textit{logarithm} of ) the $\langle \sigma_z \rangle$.  $\langle \sigma_z \rangle$ is \textit{symmetrical} (in contrast to anti symmetrical)  about $E = 0,$  (The asymmetry present at small $\lambda$ and large $|E|$ are numerical artifacts. ) The values of $\lambda$ plotted spans the range from being smaller than $|\vec{M}|$ to larger than $|\vec{M}|$. The patch of 0 $\langle \sigma_z (E) \rangle$ centered around $E=0$ for $\lambda > |\vec{M}| = 0.5$ corresponds to the bandgap opened up by large $\lambda$ where no propagating states exist. For a given value of $\lambda$, there exists a value of $|E|$ at which $\langle \sigma_z (|E|) \rangle$ peaks. Panel (b) shows the dispersion relation at $k_y=0$ at a given value of $\lambda = 0.8$. At this value of $\lambda$, $\langle \sigma_z(E) \rangle$ peaks at around $E = 1.4$. This corresponds to the energy below the vicinity of the lower horizontal dotted line in panel (b) where the tilted Dirac `cones' touch and begin to intersect with each other. 

The intersection of the two Dirac `cones'  results in the emergence of the smaller elliptical EEC curves enclosed with the larger EEC ellipses in panels (c) and (d) of the figure at the energies of the two horizontal dotted lines in panel (b).  The out of plane spin accumulation is asymmetrical across the smaller EEC ellipses so there is a net out of plane spin accumulation. These smaller EEC ellipses may be thought of as comprising the $k$ points with small values of $|\vec{k}|$ so that the in-plane spin accumulation direction is dominated by the $\vec{M}\cdot\vec{\sigma}$ term in the Hamiltonian rather than the SOI $(\vec{k}\times\vec{\sigma})\cdot\hat{z}\tau_z$ term. Due to the small $|\vec{k}|$ in the smaller ellipses, the same $\delta k_y$ caused by an electrical field and the resulting change in the SOI field direction has a far larger impact on the in-plane spin accumulation direction in the smaller EEC ellipses than in the larger ones at a given value of energy. Comparing between panels (c) and (d), the relative impact of the same $\delta k_y$ increases with decreasing $|\vec{k}|$ of the smaller EEC ellipses. The out of plane spin accumulation thus peak near the energy value at which the smaller EEC ellipses emerge where the two Dirac cones begin to intersect each other in panel (a). (There is a tradeoff between the $k$-space perimeter of the EECs over which the out of spin accumulation is integrated over, and the maximum magnitude of the spin accumulation as the smaller $k$ space ellipses decrease in size so that the peak value of $\lambda \sigma_z (E) \rangle$ occurs slightly above the energy value where the two Dirac cones intersect. )  

\section{Conclusion} 

In the first half of this work, we introduced the Murakami-Fujita potential firstly for a spin half system and then more generally for systems with real spin coupled to other discrete degrees of freedom. 

We argued that the effects of a constant electric field can, to first order in the field, be modeled by replacing the electric potential $\vec{E}\cdot\vec{r}$ by the Murakami-Fujita potential. We showed that the anomalous velocity and Cheng's extension of the Sundaram-Niu formalism can be recovered from the Heisenberg equation of motion on the MF potential, and that the result of the Kubo equation for the non-equilibrium distribution of an observable be recovered by treating the MF potential as a perturbation and then using standard time-independent non-degenerate perturbation theory.  

This formalism can be readily applied to emerging material systems of interest to spintronics with pseudospin and / or valley degrees of freedom. As an example, we applied our formalism to study the exemplary system of a three-dimensional topological insulator thin film system where the coupling between the top and bottom surfaces presents an additional discrete degree of freedom in addition to the real spin. We showed similar to the case where the inter-layer coupling is absent, that the direction of the out of plane spin accumulation due to the application of an in-plane electric field can be predicted from the direction of the torque needed to change the direction of the spin accumulation which depends on the momentum-dependent SOI field. The application of an out of plane electric field is necessary in order to break the antisymmetry of the spin accumulation.

\section{Acknowledgments} 
The authors acknowledge the Singapore National Research Foundation for support under NRF Award Nos. NRF-CRP9-2011-01 and NRF-CRP12-2013-01, and MOE under Grant No. R263000B10112.

\section{Appendix}
The starting point of Cheng and Niu's extension \cite{ChengRan} of the original Sundaram-Niu formalism \cite{ChioGoo1} to now include the time evolution of spin is to construct the Lagrangian from 
\[
	L = i \langle u| d_t u\rangle - \langle u |H|u \rangle + ...
\]
where the ... denotes the other quantities appearing in Eq. 2.18 of Ref. \onlinecite{ChioGoo1} like $\vec{k}\cdot\dot\vec{r}_c$ etc.) which do not affect the spin evolution. The $H=H_0 + H'$ that appears above consists of the unperturbed Hamiltonian $H_0$, and the perturbation $H'$. In Ref. \onlinecite{ChengRan}, the perturbation is an external magnetization. Here, we shall be concerned with a perturbing electric field modeled as $\vec{E}\cdot\vec{r}$.

We write $|u\rangle = \sum |\psi_i\rangle \eta_i$ where $|\psi_i\rangle$ are the eigenstates of $H_0$, and the $i$ is an index denoting the discrete DoFs. 
Now 
\begin{eqnarray*}
	i d_t |u\rangle &=& i (\partial_t + \dot\vec{k}\cdot\nabla_{\vec{k}}) (|\psi_i\rangle \eta_i) \\
	&=& i (\dot\vec{k}\cdot(\nabla_{\vec{k}} |\psi_i\rangle) \eta_i + \eta_i (\partial_t|\psi_i\rangle) + |\psi_i\rangle \dot{\eta}_i
\end{eqnarray*} 
so that
\begin{eqnarray*}
	L &=& i \langle u| d_t u\rangle - \langle u |H|u \rangle + ... \\
	&\approx&  i \langle u| d_t u\rangle - \langle u |(H_0 + H' \big|_{\vec{r} = 0} )|u \rangle + ... \\
	&=& i \big(\eta_i^*\dot{\eta}_i + \dot{\vec{k}}\cdot( \eta_j^* \langle \psi_j|\nabla_{\vec{k}}\psi_i\rangle \eta_i) \big) + \\ 
	&& \eta_j^* \langle \psi_j |H_0 + H' \big|_{\vec{r} = 0}  |\psi_i \rangle \eta_i  - \epsilon_i \eta_i^*\eta_i + ....
\end{eqnarray*} 

Varying with respect to $\eta_i$ and taking complex conjugate give
\begin{eqnarray*}
	\dot{\eta}_i^*  &=&   \eta_j^* \big( \dot{\vec{k}}\cdot \langle \psi_j|\nabla_{\vec{k}}\psi_i\rangle - i \langle \psi_j|H_0+H' \big|_{\vec{r} = 0} |\psi_i \rangle \big) +i\epsilon_i \eta_i^*, \\
	\dot{\eta}_i  &=&    \big( -\dot{\vec{k}}\cdot \langle \psi_i|\nabla_{\vec{k}}\psi_j\rangle + i\langle \psi_i|H_0+H' \big|_{\vec{r} = 0} |\psi_j \rangle\big)\eta_j - i\epsilon_i \eta_i
\end{eqnarray*}
so that for an operator $\sigma$ we have 
\begin{eqnarray*}
	\mathrm{d}_t \langle \sigma \rangle &=& \mathrm{d}_t \langle (\psi_i|\sigma|\psi_j\rangle \eta_i^*\eta_j) \nonumber \\ 
	&=& \sigma_{ij} \Big(  \eta_a^* \big( \dot{\vec{k}}\cdot \langle \psi_a|\nabla_{\vec{k}}\psi_i\rangle -i \langle \psi_a|H_0|\psi_i \rangle)\eta_j + \\ 
	&& \eta_i^*\big( -\dot{\vec{k}}\cdot \langle \psi_j|\nabla_{\vec{k}}\psi_b\rangle + i \langle \psi_j|H_0+H' \big|_{\vec{r} = 0} |\psi_b \rangle\big)\eta_b \\
	&&+ i(\epsilon_{i} \eta_i^*\eta_j - \epsilon_{j}\eta_i^*\eta_j) \Big) \nonumber \\
	&=& 2\mathrm{Re} (\sigma_{ij} \dot{\vec{k}} \cdot \langle \psi_a|\nabla_{\vec{k}}\psi_i\rangle \eta_a^*\eta_j) \\
	&& - i \langle [ H_0+H' \big|_{\vec{r} = 0} , \sigma] \rangle + i(\epsilon_{i}-\epsilon_{j}) \eta_i^*\eta_j.
\end{eqnarray*}
where in going from the 3rd to the 4th line we've made use of the fact that $\sigma$ being Hermitian gives $\sigma_{ij} = \sigma_{ji}^*$.

\end{document}